\documentclass[sn-mathphys,Numbered,iicol]{sn-jnl}

\usepackage{graphicx}%
\usepackage{multirow}%
\usepackage{amsmath,amssymb,amsfonts}%
\usepackage{amsthm}%
\usepackage{mathrsfs}%
\usepackage[title]{appendix}%
\usepackage{xcolor}%
\usepackage{textcomp}%
\usepackage{manyfoot}%
\usepackage{booktabs}%
\usepackage{algorithm}%
\usepackage{algorithmicx}%
\usepackage{algpseudocode}%
\usepackage{listings}%
\usepackage{setspace}

\begin{document}

\title[Article Title]{High-Resolution Laboratory Measurements of M-shell Fe EUV Line Emission using EBIT-I}


\author*[1,2]{\fnm{Alexander J.} \sur{Fairchild}}\email{fairchild8@llnl.gov}

\author[2]{\fnm{Natalie} \sur{Hell}}\email{hell1@llnl.gov}

\author[2,3]{\fnm{Peter} \sur{Beiersdorfer}}\email{beiersdorfer@berkeley.edu}

\author[2]{\fnm{Gregory V.} \sur{Brown}}\email{brown86@llnl.gov}

\author[2]{\fnm{Megan E.} \sur{Eckart}}\email{eckart2@llnl.gov}

\author[1]{\fnm{Michael} \sur{Hahn}}\email{mhahn@astro.columbia.edu}

\author[1]{\fnm{Daniel W.} \sur{Savin}}\email{dws26@columbia.edu}

\affil[1]{\orgdiv{Columbia Astrophysics Laboratory}, \orgname{Columbia University}, \orgaddress{\city{New York}, \postcode{10027}, \state{NY}, \country{USA}}}

\affil[2]{\orgdiv{Lawrence Livermore National Laboratory}, \orgaddress{\city{Livermore}, \postcode{94550}, \state{CA}, \country{USA}}}

\affil[3]{\orgdiv{Space Science Laboratory}, \orgname{University of California, Berkeley}, \orgaddress{\city{Berkeley}, \postcode{94720}, \state{CA}, \country{USA}}}


\abstract{Solar physicists routinely utilize observations of Ar-like Fe\,{\sc ix} and Cl-like Fe\,{\sc x} emission to study a variety of solar structures. However, unidentified lines exist in the Fe\,{\sc ix} and Fe\,{\sc x} spectra, greatly impeding the spectroscopic diagnostic potential of these ions. Here, we present measurements using the Lawrence Livermore National Laboratory EBIT-I electron beam ion trap in the wavelength range 238--258\,\AA{}. These studies enable us to unambiguously identify the charge state associated with each of the observed lines. This wavelength range is of particular interest because it contains the Fe\,{\sc ix} density diagnostic line ratio 241.74\,\AA{}/244.91\,\AA{}, which is predicted to be one of the best density diagnostics of the solar corona, as well as the Fe\,{\sc x} 257.26\,\AA{} magnetic-field-induced transition. We compare our measurements to the Fe\,{\sc ix} and Fe\,{\sc x} lines tabulated in CHIANTI v10.0.1, which is used for modeling the solar spectrum. In addition, we have measured previously unidentified Fe\,{\sc x} lines that will need to be added to CHIANTI and other spectroscopic databases.}

\maketitle

\section{Introduction}\label{sec1}
Accurate atomic data and line identifications for coronal ions in the extreme ultraviolet (EUV) are important for many solar missions, in particular the Extreme Ultraviolet Imaging Spectrometer (EIS) on \emph{Hinode} \cite{Culhane_2007}. EIS observes EUV emission in two spectral bands: 166–212\,\AA{} and 245–291\,\AA{}. EUV spectroscopy is essential for understanding the physics of the solar atmosphere and the interpretation of the solar spectra relies on accurate and complete plasma models. The most commonly used spectral model in solar physics is CHIANTI \cite{Delzanna_2021}. It is crucial for the atomic data included to be complete, as many emission lines are partially blended. This is especially a problem for weak emission lines, which are often used as density diagnostics \cite{Mason_1994}. Additionally, missing line identifications mean that observations are not fully utilizing all of the diagnostic information available.

Here, we present EUV spectra recorded at the Lawrence Livermore National Laboratory EBIT-I electron beam ion trap in the wavelength range between 238--258\,\AA. High-resolution line surveys for astrophysics and solar physics have been conducted using EBIT-I for several decades now and the procedure is well developed \cite{Brown_1998, Beiersdorfer_2003, Beiersdorfer_2008, Beiersdorfer_2014b, Trabert_2014a, Trabert_2014b, Beiersdorfer_2018}. Our measurements build on previously reported work in this wavelength range \cite{Trabert_2022}. This spectral region is of considerable importance for solar physics because it contains two promising plasma diagnostics. 

First, there is the potential magnetic field diagnostic using the magnetic-field-induced Fe\,{\sc x} transition at 257.26\,\AA{} \cite{Landi_2020, Xu_2022}. The magnetic field of the solar corona plays a fundamental role in solar physics, as it underpins many aspects of coronal phenomena. In general, measurements of the coronal magnetic field are challenging due to the weakness of the magnetic field signatures. This has served to motivate the increasing interest in developing a magnetic field diagnostic using the Fe\,{\sc x} 257.26\,\AA{} line, since it is one of the strongest lines in the Fe\,{\sc x} spectrum in the solar corona.

Second, there is the Fe\,{\sc ix} 241.74\,\AA{}/244.91\,\AA{} line ratio, one of the most sensitive EUV density diagnostics for the solar corona. This ratio varies in intensity by over three orders of magnitude for densities from $10^{8}$\,cm$^{-3}$ to $10^{12}$\,cm$^{-3}$. This range is to be contrasted with commonly used density sensitive coronal line ratios which are only usable over a couple of decades in density \cite{Young_2009}. Despite the remarkable density sensitivity of the Fe\,{\sc ix} 241.74\,\AA{}/244.91\,\AA{} density diagnostic, and although these lines have been measured in the Sun before \cite{Heroux_1974,Thomas_1994,Dere_1979}, this density diagnostic is not currently utilized. This is, in part, because of previous discrepancies between theory and solar observations \cite{Storey_2001}. The project reported here is part of an experimental campaign to address this discrepancy. A first step in providing this laboratory benchmark of the Fe\,{\sc ix} density diagnostic is to provide accurate line identifications and wavelength measurements, which is the subject of the present work.

\section{Experiment}\label{sec2}
The measurements were carried out at LLNL EBIT-I, which has been described in detail elsewhere \cite{Levine_1989, Beiersdorfer_2008}. EBIT-I can be used to create and trap Fe ions relevant for solar coronal physics \cite{Beiersdorfer_2014b, Beiersdorfer_2018, Arthanayaka_2020, Trabert_2022}.

In the trapping region, Fe ions are confined radially by the potential of the magnetically compressed electron beam and axially by voltages applied to the outer drift tubes. The magnetic field in the trap is estimated to be 3\,T. The axial voltages also set the electron beam energy at a nominal electron beam energy, $E_\mathrm{e}$. The Fe ions are produced by injecting iron pentacarbonyl [Fe(CO)$_5$] into the trapping region. The molecules are rapidly dissociated and ionized by electron impact. The ionization states of the Fe ions present in the trap are largely determined by the energy of the electron beam. The maximum energy of the electron beam energy determines the maximum ionization stage in the trap region. The electron beam energy is nearly monoenergetic and lower charge states exist primarily because of the continuous injection of Fe(CO)$_5$. EBIT-I operates at typical electron densities $\leq  5 \times 10^{11}$\,cm$^{-3}$, which are relevant for studying the solar corona.

The EUV spectra were recorded using the high-resolution grazing-incidence grating spectrometer (HiGGS) \cite{Beiersdorfer_2014a}. At EUV wavelengths, for any given setting it observes an $\sim 20$\,\AA{} wide bandpass. The spectra reported here have been calibrated using the NIST Atomic Spectra Database Ritz wavelengths for strong O IV, N IV, N V, and Fe XIV lines \cite{NIST_ASD}. The calibration residuals are everywhere less than 5\,m\AA{}. The weighted root mean square of the calibration residuals is 1.5\,m\AA{} and has been added in quadrature with the statistical wavelength uncertainties from line fitting, to give the full measurement uncertainties. The resulting quadrature sums are reported in Table \ref{table:wvl}.

\section{Results}\label{sec3}
\begin{figure*}[ht!]
\centering
\includegraphics[width=0.9\textwidth]{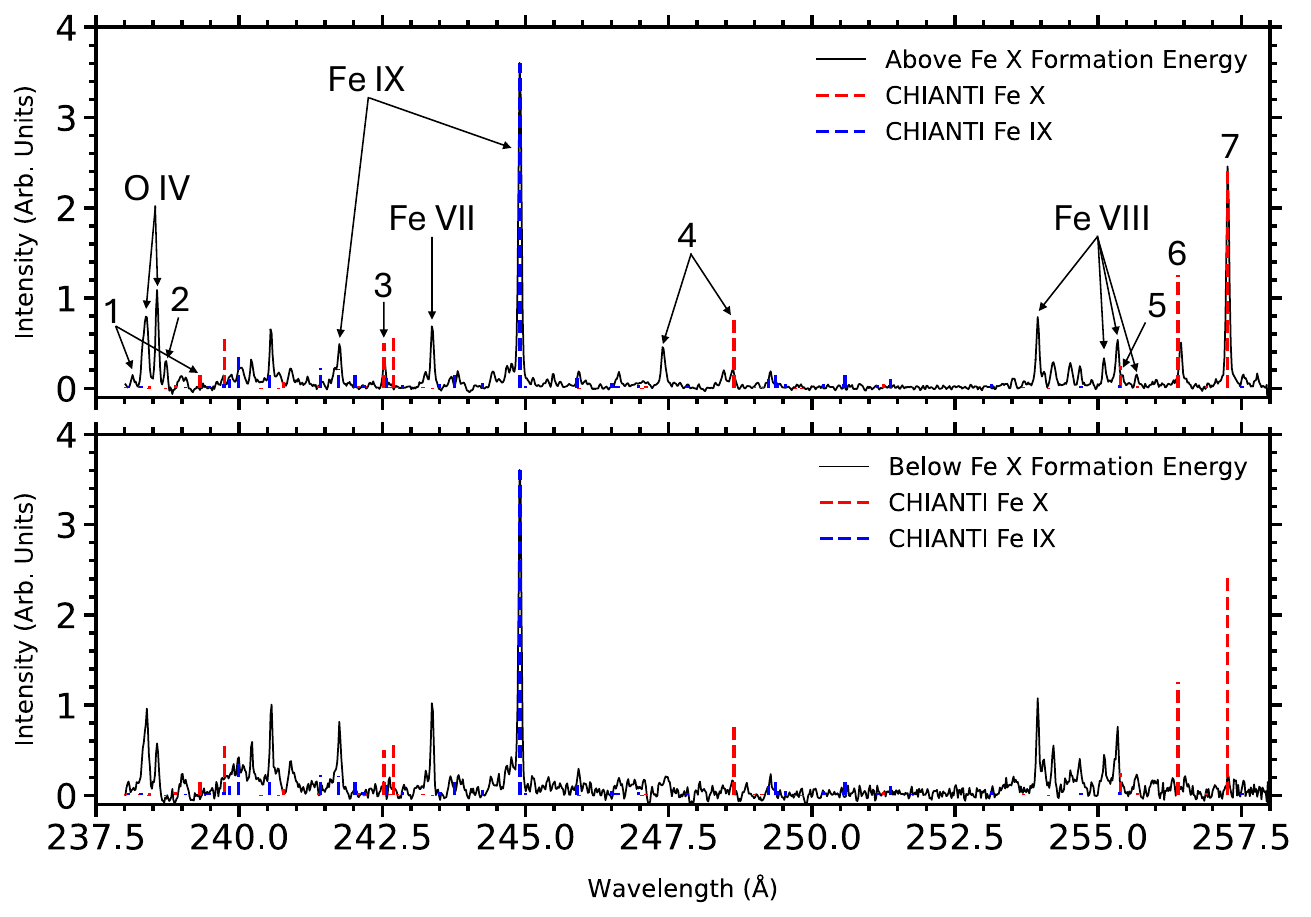}
\caption{EBIT-1 EUV spectra measured using the HiGGS in the 238--258\,\AA{} bandpass for electron beam energies above (top) and below (bottom) the Fe\,{\sc ix} ionization potential, i.e., the Fe\,{\sc x} formation energy. The red (blue) dashed lines provide the location and relative intensity of the most prominent Fe\,{\sc x} (Fe\,{\sc ix}) lines tabulated in CHIANTI v10.0.1 for a plasma at log[T(K)] = 6.0 and density of $1 \times 10^{11}$\,cm$^{-3}$. Selected lines are labeled by their isoelectronic sequence. The Fe\,{\sc x} lines are numbered according to the key in Table \ref{table:wvl}.}
\label{fig:fex}
\end{figure*}

\begin{table*}{}
\centering
\resizebox{\textwidth}{!}{%
\begin{tabular}{p{0.8em}ccccccccc}
\hline
Key & Transition & Experiment & Line Intensity & CHIANTI & MCDHF & $\Delta$(E-C) & $\Delta$(E-M) & $\Delta$(C-M) \\
\hline
\hline

1 & ${}^2$P$_{1/2}$  $\rightarrow$ ${}^2$P$_{3/2}$ & 238.144(7) & 11(2) & 239.314 & 238.20 & -1.17 & -0.056 & 1.114 \\
2 & $-$ & 238.72(3) & 22(2) & - & - & - & - & - \\
3 & ${}^2$P$_{3/2}$ $\rightarrow$ ${}^2$P$_{1/2}$ & 242.536(4) & 23(2) & 242.531 & 242.59 & 0.005 & -0.054 & -0.059 \\
4 & ${}^2$P$_{1/2}$ $\rightarrow$ ${}^2$P$_{1/2}$ & 247.406(2) & 38(2) & 248.646 & 247.47 & -1.24 & -0.064 & 1.176 \\
5 & ${}^4$D$_{1/2}$ $\rightarrow$ ${}^2$P$_{3/2}$ & 255.433(9) & 10(2) & 255.393 & 255.53 & 0.040 & -0.097 & -0.137 \\
6 & ${}^4$D$_{3/2}$ $\rightarrow$ ${}^2$P$_{3/2}$ & 256.441(3) & 38(2) & 256.398 & 256.55 & 0.043 & -0.109 & -0.152 \\
7 & ${}^4$D$_{5/2}$ $\rightarrow$ ${}^2$P$_{3/2}$ & 257.265(2) & 175(2) & 257.259 & 257.36 & 0.006 & -0.095 & -0.101 \\

\hline
\end{tabular}
}
\caption{Measured Fe\,{\sc x} wavelengths compared with the tabulated values in CHIANTI v10.0.1 \cite{Delzanna_2021} and the multiconfiguration Dirac-Hartree-Fock (MCDHF) calculations of Wang et al. \cite{Wang_2020}. Note - The key indicates the Fe\,{\sc x} line as it is numbered in Figure \ref{fig:fex}. The configuration for each transition is $3s^23p^43d$ $\rightarrow$ $3s^23p^5$ and so we indicate a given transition using only the terms and levels involved. The ground level is $3s^23p^5$ ${}^2$P$_{3/2}$. The numbers in parentheses after the measured quantities are the uncertainties in the last digit. $\Delta$(E-C) is the difference between the EBIT-I measured wavelength and the wavelength listed in CHIANTI. $\Delta$(E-M) is the difference between the EBIT-I measured wavelength and the MCDHF calculated wavelength. $\Delta$(C-M) is the difference in the wavelengths between CHIANTI and the MCDHF calculation. All wavelengths and wavelength differences are given in units of \AA{}.}
\label{table:wvl}
\end{table*}

\begin{table*}{}
\centering
\begin{tabular}{cccccccc} 
\hline
Level & Experiment & CHIANTI & MCDHF & $\Delta$(E-C) & $\Delta$(E-M) & $\Delta$(C-M) \\
\hline
\hline

$3s^23p^43d\ {}^4$D$_{5/2}$ & 388704(2) & 388713.5 & 388558 & -10 & 146 & 156 \\
$3s^23p^43d\ {}^4$D$_{3/2}$ & 389953(4) & 390019 & 389791 & -66 & 162 & 228 \\
$3s^23p^43d\ {}^4$D$_{1/2}$ & 391492(13) & 391554 & 391341 & -62 & 151 & 213 \\
$3s^23p^43d\ {}^2$P$_{1/2}$ & 419914(12) & 417861 & 419812 & 2053 & 102 & -1951 \\
$3s^23p^43d\ {}^2$P$_{3/2}$ & 428030(14) & 428002 & 427951 & 28 & 79 & 51 \\

\hline
\end{tabular}
\caption{Measured Fe\,{\sc x} energy levels compared with the tabulated values in CHIANTI v10.0.1 \cite{Delzanna_2021} and the Multiconfiguration Dirac-Hartree-Fock (MCDHF) calculations of Wang et al. \cite{Wang_2020}. Note - The CHIANTI energy levels are tabulated measured values except for the $3s^23p^43d\ {}^2$P$_{1/2}$, which has no measured value. The numbers in parentheses after the measured quantities are the uncertainties in the last digit(s). $\Delta$(E-C) is the difference between the EBIT-I measured energy level and the energy level listed in CHIANTI. $\Delta$(E-M) is the difference between the EBIT-I measured energy level and the MCDHF calculated energy level. $\Delta$(C-M) is the difference in the energy levels between CHIANTI and the MCDHF calculation. All energy levels and energy level differences are presented in units of cm$^{-1}$}
\label{table:energy}
\end{table*}

In order to unambiguously identify important Fe lines and separate possible line blends in EUV bandpasses relevant for solar observations, we are carrying out a series of line survey measurements. These studies build on the previous 30 years of EBIT-I work in the EUV \cite{Beiersdorfer_1999, Lepson_2002, Beiersdorfer_2012, Trabert_2014b, Trabert_2014a, Beiersdorfer_2014b, Beiersdorfer_2018, beiersdorfer_2022, Lepson_2023}. Figure \ref{fig:fex} shows the EBIT-I spectra recorded in the 238--258\,\AA{} range at electron beam energies above and below the energy required to form Fe\,{\sc x}, i.e., the energy of the Fe\,{\sc ix} ionization potential. Emission lines from a given charge state appear when the beam energy reaches the ionization threshold to form that charge state. By measuring the spectrum while varying the beam energy in steps, the various emission lines can be unambiguously associated with their emitting charge state. In Fig. \ref{fig:fex}, the nominal electron beam energies are $E_\mathrm{e} = 230$\,eV (top) and $E_\mathrm{e} = 170$\,eV (bottom). The solid black line is the EBIT-I spectrum and the dashed vertical lines denote the location of the most prominent Fe\,{\sc x} and Fe\,{\sc ix} lines in this wavelength range, as tabulated in the CHIANTI v10.0.1 database \cite{Delzanna_2021}. 

The top panel of Fig. \ref{fig:fex} contains prominent emission lines from Fe\,{\sc vii}--{\sc x}. We identify the Fe\,{\sc x} lines in the top panel by observing which lines are absent in the bottom panel. The measured Fe\,{\sc x} line positions are listed in Table \ref{table:wvl} and compared to the wavelengths tabulated in CHIANTI v10.0.1 \cite{Delzanna_2021} and the recent multiconfiguration Dirac-Hartree-Fock (MCDHF) calculations of Wang et al. \cite{Wang_2020}. The Fe\,{\sc x} lines listed in Table \ref{table:wvl} and all other Fe\,{\sc x} transitions discussed in this manuscript, are $3s^23p^43d$ $\rightarrow$ $3s^23p^5$ configurations and so we will refer to a given transition by its term and level. We note that the ground level is $3s^23p^5$ ${}^2$P$_{3/2}$. 

The measured and tabulated wavelengths in CHIANTI for the ${}^2$P$_{3/2}$ $\rightarrow$ ${}^2$P$_{1/2}$ and the ${}^4$D$_{5/2}$ $\rightarrow$ ${}^2$P$_{3/2}$ transitions show good agreement on the level of 5-6\,m\AA{}. However, the ${}^4$D$_{1/2}$  $\rightarrow$ ${}^2$P$_{3/2}$ and ${}^4$D$_{3/2}$ $\rightarrow$ ${}^2$P$_{3/2}$ transitions disagree on the order of 40\,m\AA{}. This disagreement is much larger than the wavelength uncertainty in the experiment, which ranges between 2--9\,m\AA{}. We suggest that the CHIANTI wavelengths will need to be updated. In addition, we propose that the ${}^2$P$_{1/2}$ $\rightarrow$ ${}^2$P$_{1/2}$ line (line 4 in Fig.~\ref{fig:fex}) at 247.406\,\AA{} in the EBIT-I spectrum corresponds to the unidentified line at 247.404\,\AA{} previously observed using EIS \cite{DelZanna_2012}.

The ${}^2$P$_{1/2}$ $\rightarrow$ ${}^2$P$_{3/2}$ and ${}^2$P$_{1/2}$ transitions (lines 1 and 4) are flagged as ``unobserved'' in CHIANTI. The EBIT-I measured wavelengths for these two transitions are 238.144\,\AA{} and 247.406\,\AA{}, respectively. In CHIANTI when a transition is flagged as ``unobserved'' this means there is not an experimental value for one or both of the energy levels. Using either of the EBIT-I wavelengths to determine the upper level ${}^2$P$_{1/2}$ energy gives the same value within the stated uncertainties. This gives confidence in our transition identifications.

The 247.406\,\AA{} line might also be of interest for magnetic field diagnostic line ratios involving the Fe\,{\sc x} 257.26\,\AA{} line, which consists of the magnetic quadrupole (M2) ${}^4$D$_{7/2} \rightarrow {}^2$P$_{3/2}$ and the electric dipole (E1) ${}^4$D$_{5/2} \rightarrow {}^2$P$_{3/2}$ transitions. In the presence of a magnetic field, the ${}^4$D$_{7/2}$ level mixes with the ${}^4$D$_{5/2}$ level, making the line ratio sensitive to the magnetic field strength. Unfortunately, these transitions are too close in wavelength to spectrally resolve. Hence, recent laboratory measurements have instead used the E1 Fe\,{\sc x} 226.31\,\AA{} line as a normalization line \cite{Xu_2022}. Our identification of other Fe\,{\sc x} E1 lines, closer in wavelength to the Fe\,{\sc x} 257.26\,\AA{} line, may offer better candidates for normalization lines for the magnetic field diagnostic. This is especially true if the ${}^2$P$_{1/2}$ $\rightarrow$ ${}^2$P$_{1/2}$ transition has indeed been observed using EIS. We note that in Table \ref{table:wvl}, for the Fe\,{\sc x} 257.265\,\AA{} line, we only list the E1 ${}^4$D$_{5/2}$ $\rightarrow$ ${}^2$P$_{3/2}$ transition because we expect negligible contribution from the M2 ${}^4$D$_{7/2}$ $\rightarrow$ ${}^2$P$_{3/2}$ due to collisional quenching at the electron density in EBIT-I.

CHIANTI lists three additional ``unobserved'' transitions in this wavelength range which decay to $3s^23p^5$\,${}^2$P$_{1/2}$, the first excited level: ${}^2$D$_{3/2}$, ${}^4$P$_{1/2}$, and ${}^4$F$_{3/2}$ $\rightarrow$ ${}^2$P$_{1/2}$. The CHIANTI wavelengths of these transitions are 239.742, 240.778, 242.692\,\AA{}, respectively. In principle, any of these three transitions could correspond to the unidentified line in Table \ref{table:wvl} at 238.72\,\AA{} (line 2). Unfortunately, the transitions for all of these upper levels to the ground level fall outside the wavelength range studied here, preventing a definitive identification of the transition. It is worth noting that the MCDHF calculations of Wang et al. show that the ${}^4$F$_{3/2}$ $\rightarrow$ ${}^2$P$_{3/2}$ transition has a branching fraction close to 0.99, where the branching fraction considers radiative decay to all lower levels. This means that the ${}^4$F$_{3/2}$ $\rightarrow$ ${}^2$P$_{1/2}$ transition most likely would not show up in the EBIT-I spectrum. In contrast, CHIANTI has a much lower branching fraction close to 0.27, although CHIANTI only lists decay rates from the ${}^4$F$_{3/2}$ upper level to the two lowest lying levels ${}^2$P$_{1/2}$ and ${}^2$P$_{3/2}$.

In Table \ref{table:energy}, we present the measured upper level energies of the transitions listed in Table \ref{table:wvl}, and compare these energy levels with those tabulated in CHIANTI and with the MCDHF calculations of Wang et al. \cite{Wang_2020}. For the transitions to the ground level (lines 1, 5, 6, and 7), the measured transition wavelengths directly correspond to the energy of the upper level. To determine the $3s^23p^43d$\,${}^2$P$_{3/2}$ upper level energy, we used the wavelengths of the $3s^23p^43d$\,${}^2$P$_{3/2}$ $\rightarrow$ $3s^23p^5$\,${}^2$P$_{1/2}$ (line 3) and the $3s^23p^43d$\,${}^2$P$_{1/2}$ $\rightarrow$ $3s^23p^5$\,${}^2$P$_{1/2}$ (line 4) transitions together with our measured value for the $3s^23p^43d$\,${}^2$P$_{1/2}$ level (from line 1). Since both transitions have the same lower level, the difference in their wavelengths corresponds to the difference in their upper level energies. By adding this measured energy difference with our measurement of the $3s^23p^43d$\,${}^2$P$_{1/2}$ level we were able to determine the $3s^23p^43d$\,${}^2$P$_{3/2}$ level energy. Lastly, we note that the calculated value for the $3s^23p^43d$\,${}^2$P$_{1/2}$ level provided by Wang et al. is much closer to the EBIT-I measurement than the theoretical value currently tabulated in CHIANTI.

\section{Conclusions}\label{sec4}
We have compared our EBIT-I EUV measurements in the 238-258\,\AA{} wavelength range with the tabulated wavelengths and energy levels in the CHIANTI database and recent MCDHF calculations for Fe\,{\sc x}.  We have measured two previously unidentified transitions from the same upper level $3s^23p^43d\ {}^2$P$_{1/2}$ to the levels $3s^23p^5\ {}^2$P$_{1/2}$ and $3s^23p^5\ {}^2$P$_{3/2}$. In addition, we have measured the wavelength of an unidentified Fe\,{\sc x} line at 238.72(3)\,\AA{}. However, more experiments are required to determine precisely the upper and lower levels in this transition. The comparisons between the EBIT-I measurements, CHIANTI, and the MCDHF calculations suggest that in absence of experimental values for the Fe\,{\sc x} energy levels the MCDHF calculations should be preferred.

This work is still ongoing. Future studies will focus on completing the line surveys for the additional Fe charge states. We will also systemically investigate the possibility of line blending affecting either line in the Fe\,{\sc ix} line ratio 241.74\,\AA{}/244.91\,\AA{} and determine the Fe\,{\sc ix} line intensities as a function of the effective electron density in EBIT-I.

\backmatter

\bmhead{Acknowledgments}
This work is supported, in part, by NASA H-TIDeS grant NNX16AF10G. Work at Lawrence Livermore National Laboratory is performed under the auspices of the U. S. Department of Energy under contract No. DE-AC52-07NA27344. We are grateful to the anonymous referees for their comments and suggestions, which greatly improved this manuscript.

\bmhead{Author Contributions}
Conceptualization and project administration: PB, GVB, MH, and DWS; Methodology: PB, GVB, MEE, NH, AJF, MH, and DWS; Investigation: PB, AJF, and NH; Formal analysis: PB, AJF, NH, MH, and DWS; Software: AJF, and NH; Writing---original draft: AJF; Writing---review \& editing: PB, GVB, MH, NH, and DWS.

\bmhead{Data Availability Statement}
The manuscript has associated data in a data repository. [Author’s comment: Data sets generated during the current study are available from the corresponding author on reasonable request. The manuscript has associated data in a data repository.]

\bibliography{mnemonic,sn-bibliography}

\end{document}